# Estimation of Total Fusion Reactivity and Contribution from Supra-thermal Tail using 3-parameter Dagum Ion Speed Distribution

Rudrodip Majumdar[1*], Debraj Das[2]

[1] North Carolina State University, Department of Nuclear Engineering,
  Raleigh, NC 27695 USA.

[2] North Carolina State University, Department of Statistics,
Raleigh, NC 27695-8203 USA.

*rmajumd@ncsu.edu* , *ddas3@ncsu.edu*

**Abstract.** Thermonuclear fusion reactivity is a pivotal quantity in the studies pertaining to fusion energy production, fusion ignition and energy break-even analysis in both inertially and magnetically confined systems. Although nuclear fusion reactivity and thereafter the power density of a magnetic confinement fusion reactor and the fulfillment of the ignition criterion are quantitatively determined by assuming the ion speed distribution to be Maxwellian, a significant population of suprathermal ions, with energy greater than the quasi-Maxwellian background plasma temperature, is generated by the fusion reactions and auxiliary heating in the fusion devices. In the current work 3-parameter Dagum speed distribution has been introduced to include the effect of suprathermal ion population in the calculation of total fusion reactivity. The extent of enhancement in the fusion reactivity, at different back-ground temperatures of the fusion fuel plasma, due to the suprathermal ion population has also been discussed.

Rudrodip Majumdar *
North Carolina State University, Department of Nuclear Engineering, Raleigh, NC 27695 USA ,  e-mail: rmajumd@ncsu.edu

* Corresponding author

Debraj Das
North Carolina State University, Department of Statistics, Raleigh, NC 27695-8203, USA.
e-mail: *ddas3@ncsu.edu*

**Introduction**

A mixture containing a considerably high number density ($10^{20}$-$10^{21}$ particles/m$^3$) of certain light elements can be made to fuse if the temperature of the mixture is raised to a sufficiently high value (~ few keV). The most likely fuels for near-term magnetically confined fusion devices are the isotopes of hydrogen and helium. Particularly the reaction that occurs in a 1:1 mixture of the hydrogen isotopes deuterium ($_1^2H$ or D) and tritium ($_1^3H$ or T) is a popular choice because the D-T cycle is easiest to achieve, and it has the higher reaction rate compared to the other candidate reactions **[1, 2]**. Power-balance calculations and energy break-even analysis shows that it is more difficult to achieve the ignition condition in a D-T mixture than to meet the Lawson criterion **[2]**. Thus, it is reasonable to include the alpha-particle heating of the plasma and energy externally supplied to the plasma with a view to balancing the energy losses, in the power-balance calculations. The energy of charged fusion products (e.g. alpha particles) on being diverted directly to fuel-ions, gives rise to non-Maxwellian fuel ion distributions and temperature differences between the interacting species. Additionally, a temperature difference between electrons and ions is also observed in such a case. For D-T and D-$^3$He reactors, with 75% of charged fusion product power being diverted to fuel ions, temperature differences between electrons and ions increase the total fusion reactivity by 40-70%, while non-Maxwellian fuel ion distributions and temperature differences between ionic species are capable of enhancing the reactivity by an additional 3-15%, all the enhancements being calculated relative to the isothermal Maxwellian case where the ion and electron kinetic temperatures are assumed to be equal **[3]**. Computational results related to the effect of unequal ion temperatures on the total fusion reactivity have been reported previously and for D-T reactions the results showed that it is always advantageous to have hotter deuterons than tritons for the same average kinetic temperature of the two species **[4]**. Primarily, the interaction of the energetic charged particles produced in exothermic nuclear reactions with the fuel plasma enhances the tail part of the Maxwellian spectrum **[4, 5]**. Previously, enhanced tail of ion speed distribution has been modeled mathematically using either the tail-enhanced variants of the generalized Maxwellian distribution **[4, 6],** or the Lorentzian (Kappa) speed distribution **[4, 7]** that reasonably describes a power-law suprathermal tail comprising of accelerated hot ion population. Effect of anisotropy on the fusion reactivity arising from the ion drifts have been discussed previously using drifting tri-Maxwellian ion velocity distribution **[8]**. The current interest is to investigate the 3-parameter Dagum Distribution in the context of analyzing the fractional contribution of the fusion reactivity from supra-thermal tail of the ion speed distribution. The Generalized Maxwellian distribution has an exponentially decaying tail, while in our case a polynomially decaying pronounced suprathermal tail was of interest. There are many distributions with polynomially decaying tails. The primary motivation behind adapting the Dagum distribution in particular besides the prevalent use of Maxwellian and Kappa Speed Distribution is that a special case of the Dagum distribution can be seen as a 'continuous mixture' of Maxwell distribution. It is quite well-known that the Maxwellian distribution is a special case of a 'generalized gamma' (GG) distribution, and the Dagum distribution can be obtained as a compound generalized gamma distribution whose scale parameter follows an inverse Weibull distribution **[9]**. Additionally the Kappa distribution can be reduced to Dagum distribution with a change parametric treatment **[9].** Thus the transition from Maxwellian and Kappa distributions to one of its relatives was quite natural and not far-fetched in the context of fusion reactivity analysis. Dagum distribution contains shape parameters which provides us with flexibility and enables us in describing theoretically the various non-thermalized scenarios in the fusion plasma that deviates considerably from quasi-Maxwellian state.

**Brief description to 3-Parameter Dagum Speed Distribution**

Dagum distribution was introduced in 1970s by statistician and economist Camilo Dagum **[10, 11]**. The density function of 3-parameter Dagum speed distribution in center of mass (C.O.M) frame of reference is defined as follows:

$$f_{Dg}(v) = \frac{ap}{v}\left(\frac{\left(\frac{v}{b_{Dg}}\right)^{ap}}{\left(\left(\frac{v}{b_{Dg}}\right)^a + 1\right)^{p+1}}\right) I_v(0,\infty), \quad (a,p,b > 0) \tag{1}$$

where, $I_x(A) = \begin{cases} 1, & \text{if } x \in A \\ 0, & \text{if } x \notin A \end{cases}$. Here, $v$ is the relative speed between the two interacting species Deuteron and Triton. The parameters $a$ and $p$ are shape parameters and $b_{Dg}$ is the scale parameter. The scale parameter $b_{Dg}$ is defined as follows:

$$b_{Dg} = \sqrt{\frac{k_B T_{Dg}}{\mu}}$$

where, $k_B$ is the Boltzmann constant, $T_{Dg}$ is the background bulk temperature of the ion population following the speed distribution, $\mu$ is the reduced mass of a two body system comprising of Deuteron & Triton, assuming Triton at rest. In the current work, the range of the shape parameter $(a, p)$ has been represented in the form of the set $S = \{(a,p) \in \Re^2 : ap > 1 \text{ and } a > 2\}$. Choosing $a > 2$ guarantees a finite variance of the speed distribution. Additionally, by assuming $ap > 1$, the existence of the interior mode in the speed distribution, $v_{mode} = b_{Dg}\left(\frac{ap-1}{a+1}\right)^{1/a}$, is ensured; since a distinct peak is an important characteristic of an ion speed distribution of interest. Figure 1 presents Dagum speed distributions for a fixed value of the shape parameter 'a' (a = 3.0) and four different chosen values of parameter 'p' (p= 5.0, 6.0, 7.0, 8.0), for an interacting D-T system at an average kinetic temperature of 15 keV.

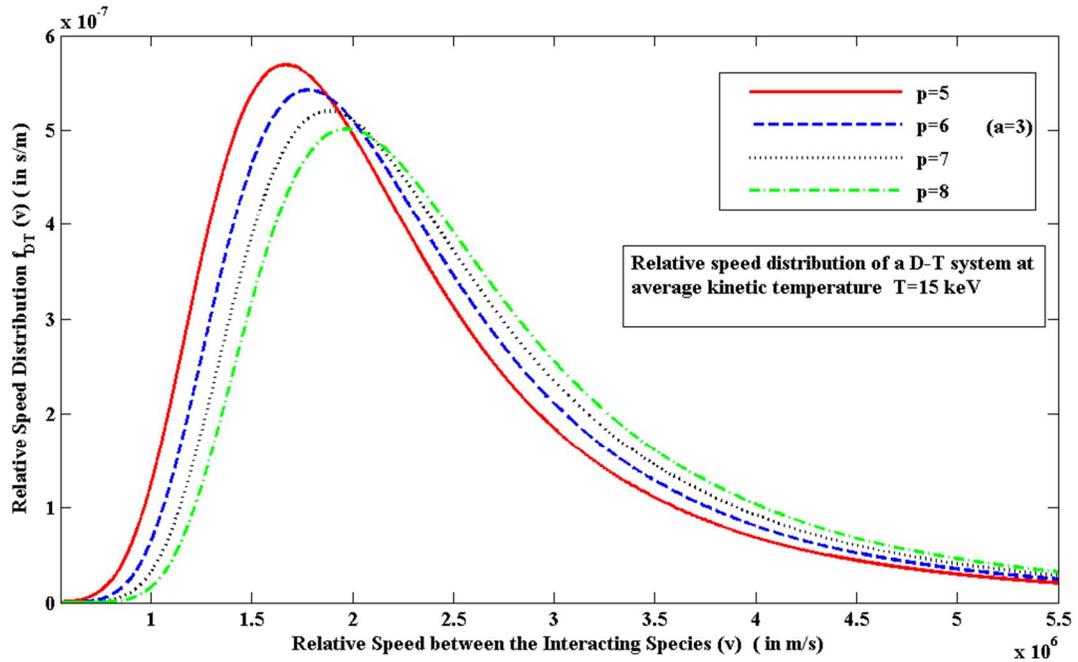

Fig. 1 Dagum speed distribution vs ion speed (C.O.M) plot for a fixed value of 'a', and multiple values of 'p'

For a reacting D-T system, Coulomb collisions combined with power balance requirements result in an optimum operating temperature of the order of 15 keV **[1]** and that prompted the choice of 15 keV as a standard kinetic temperature. Calculations show that for a 15 keV D-T fusion plasma with an energy confinement time of 1 s, a pressure of about 8 atm is required for the plasma to be ignited; and thereafter it is sustained purely by the self-heating of the fusion alpha particles **[1]**, which may eventually lead to the formation of a non-thermalized suprathermal population. Figure 2 presents Dagum speed distributions for a fixed value of the shape parameter 'p' (p= 5.0) and three different chosen values of parameter 'a' (a= 2.5, 3.5, 4.5) for the same system at the same average kinetic temperature.

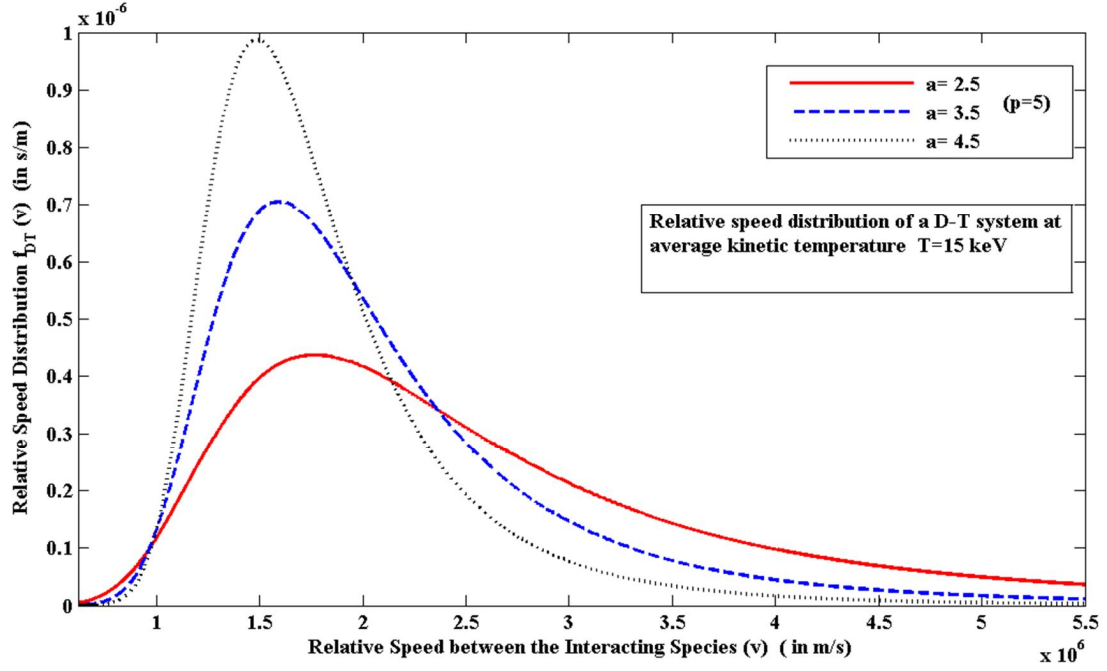

Fig. 2 Dagum speed distribution vs ion speed (C.O.M) plot for a fixed value of 'p', and multiple values of 'a'

Looking at the Figures 1 and 2, it can be clearly seen that the ion speed distribution is affected more by the change in the value of parameter 'a', compared to that in case of parameter 'p'. One of the most important reasons behind considering Dagum distribution in describing the nature of the non-thermalized accelerated ions is that there is a natural ordering within the family of this distribution with respect to the thickness of the tail based on the shape parameters. It is well-known that a random variable $X$ is stochastically greater than another random variable $Y$, sometimes denoted by $X \geq_{st} Y$; if $P(X > x) \geq P(Y > x)$ for all real number $x$. Hence, for two non-negative random variables $X$ & $Y$, $X \geq_{st} Y$ means that the distribution of $Y$ has a thinner tail compared to that of $X$. Characterization of stochastic ordering within the class of all Dagum distributions in terms of the associated parameters had been established by Klonner **[12]**. Mathematically, the characterization is expressed as follows:

If $X_i$ follows $Dg(a_i, p_i, b_i)$, $i = 1, 2$ then,

$$X \geq_{st} Y \quad if \ a_1 \leq a_2, \ a_1 p_1 \geq a_2 p_2 \ and \ b_1 \leq b_2 .$$

So, $Dg(a_1, p, b)$ has more pronounced tail than $Dg(a_2, p, b)$ if $a_1 \leq a_2$ and $Dg(a, p_1, b)$ has more pronounced tail than $Dg(a, p_2, b)$ if $p_1 \geq p_2$. The shapes of the speed distribution, together with the energy-dependent fusion cross-sections determine the total reactivities.

Besides looking into the effect of changes in the Dagum shaping parameter values on the relative speed distribution, another interest is to look into the effect of change in the average kinetic temperature on the speed distribution for fixed values of 'a' and 'p'. Figure 3 presents the Dagum speed distributions for average ion kinetic temperatures of 15, 40, 60 and 80 keV respectively, for a=3.0 and p= 5.0; and as can be seen from the figure, as the average ion kinetic temperature increases, the Dagum distribution flattens out considerably while exhibiting a more pronounced tail.

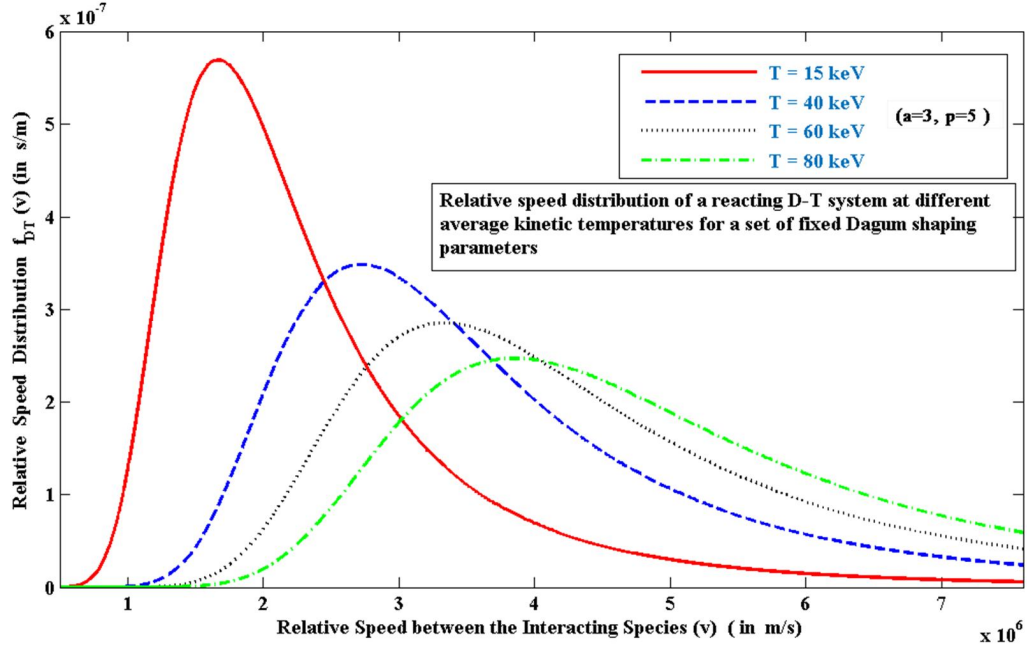

Fig. 3 Dagum speed distribution vs ion speed (C.O.M) plot for fixed value of 'a' and 'p', and different values of average ion kinetic temperatures

**Results and Discussions**

Thermonuclear fusion reactivity pertaining to a particular fusion reaction can be calculated using the following mathematical expression, well-known from the published literature **[4, 5, 8]**:

$$\langle \sigma v \rangle = \int_0^\infty v\, f(v) \sigma(v) dv \tag{2}$$

In the present case, $v$ represents the relative speed between the interacting species in Centre-of-Mass frame of reference, $f(v)$ represents the relative speed distribution and $\sigma(v)$ represents the microscopic fusion cross section for a particular fusion reaction as a function of the Centre-of-Mass energy for that reaction system. In the current work, improved fusion cross-section formulae derived using R-Matrix theory, as prescribed by Bosch and Hale, have been used to calculate the reactivities [13]. Fusion reactivity thus obtained is used directly to calculate the fusion power density [1].

## A. Comparison of Total Fusion Reactivities

The current interest is to calculate the total fusion reactivities for the D-T reaction occurring at different average ion kinetic temperatures. Reactivities have been calculated for the Dagum and Maxwellian ion speed distributions and they are represented by $\langle\sigma v\rangle_{DT,Dg}$ and $\langle\sigma v\rangle_{DT,Mx}$ respectively. In the current work, it has been assumed that both of the interacting ionic species follow the same speed distribution and thus the system can be reduced to a relative speed frame of reference. Here, two different approaches have been followed for the comparison between the total reactivities obtained using the Dagum and Maxwellian speed distributions. The first approach assumes that the average ion kinetic temperature or the background temperature of the bulk plasma is the same for both the ion speed distributions of interest. On the other hand, by following the second approach, equivalent Maxwellian temperatures ($T_{Mx}$) are calculated for the chosen values of Dagum kinetic temperatures, where the equivalence is obtained by equating the respective mean thermal speeds coming from two ion speed distributions to each other. Upon obtaining all the kinetic temperatures of interest, the fusion reactivities are calculated and are compared against each other.

### a. Comparison based on same Bulk Kinetic Temperature

With a view to estimating the fusion reactivities for the aforesaid ion speed distributions in a reacting D-T system assuming the background bulk kinetic temperature to be the same in the cases of both the distributions, a wide range of plasma kinetic temperatures have been chosen. In the current work the kinetic temperature ranges from a below- breakeven temperature of about 10 keV to the very high kinetic temperature of 100 keV, a value when interpreted in terms of the projectile energy, corresponds to the peak value in the D-T fusion cross section profile [1, 14]. Temperatures of 10, 15, 25, 40, 60, 80 and 100 keV have been chosen for the comparison of the numerically estimated values of the total fusion reactivities.

Tables 1 and 2 present total fusion reactivities calculated using the Maxwellian speed distribution as well as the Dagum speed distribution for the aforesaid seven different kinetic temperatures. Table 1 exhibits the reactivities for a fixed value of Dagum shaping parameter 'a' (a= 3.0) and multiple values of parameter 'p' (p= 5.0, 6.0, 7.0 and 8.0); whereas Table 2 shows total reactivities for a fixed value of shaping parameter 'p' (p= 5.0) and multiple values of parameter 'a' (a= 2.5, 3.5, 4.0 and 4.5).

**Table 1:** $\langle\sigma v\rangle_{DT,Dg}$ ($a = 3.0$ and $p = 5.0, 6.0, 7.0, 8.0$) and $\langle\sigma v\rangle_{DT,Mx}$ for same background plasma kinetic temperature

| $T_{Dg}$ (keV) | $T_{Mx}$ (keV) | $\langle\sigma v\rangle_{DT,Dg}$ $*10^{-22}$ $(m^3/s)$ (p=5.0) | $\langle\sigma v\rangle_{DT,Dg}$ $*10^{-22}$ $(m^3/s)$ (p=6.0) | $\langle\sigma v\rangle_{DT,Dg}$ $*10^{-22}$ $(m^3/s)$ (p=7.0) | $\langle\sigma v\rangle_{DT,Dg}$ $*10^{-22}$ $(m^3/s)$ (p=8.0) | $\langle\sigma v\rangle_{DT,Mx}$ $*10^{-22}$ $(m^3/s)$ |
|---|---|---|---|---|---|---|
| 10 | 10 | 3.024 | 3.504 | 3.951 | 4.368 | 1.161 |
| 15 | 15 | 4.734 | 5.372 | 5.938 | 6.441 | 2.791 |
| 25 | 25 | 7.327 | 7.989 | 8.508 | 8.909 | 5.738 |
| 40 | 40 | 9.173 | 9.493 | 9.362 | 9.648 | 8.060 |
| 60 | 60 | 9.364 | 9.155 | 8.833 | 8.457 | 8.898 |
| 80 | 80 | 8.527 | 7.978 | 7.412 | 6.869 | 8.745 |
| 100 | 100 | 7.453 | 6.736 | 6.080 | 5.499 | 8.258 |

**Table 2:** $\langle\sigma v\rangle_{DT,Dg}$ ($a = 2.5, 3.5, 4.0, 4.5$ and $p = 5.0$) and $\langle\sigma v\rangle_{DT,Mx}$ for same background plasma kinetic temperature

| $T_{Dg}$ (keV) | $T_{Mx}$ (keV) | $\langle\sigma v\rangle_{DT,Dg}$ $*10^{-22}$ $(m^3/s)$ (a=2.5) | $\langle\sigma v\rangle_{DT,Dg}$ $*10^{-22}$ $(m^3/s)$ (a=3.5) | $\langle\sigma v\rangle_{DT,Dg}$ $*10^{-22}$ $(m^3/s)$ (a=4.0) | $\langle\sigma v\rangle_{DT,Dg}$ $*10^{-22}$ $(m^3/s)$ (a=4.5) | $\langle\sigma v\rangle_{DT,Mx}$ $*10^{-22}$ $(m^3/s)$ |
|---|---|---|---|---|---|---|
| 10 | 10 | 4.106 | 2.124 | 1.459 | 0.994 | 1.161 |
| 15 | 15 | 5.671 | 3.745 | 2.877 | 2.181 | 2.791 |
| 25 | 25 | 7.556 | 6.697 | 5.905 | 5.097 | 5.738 |
| 40 | 40 | 8.419 | 9.419 | 9.303 | 8.955 | 8.060 |
| 60 | 60 | 7.989 | 10.363 | 11.037 | 11.448 | 8.898 |
| 80 | 80 | 7.039 | 9.769 | 10.777 | 11.580 | 8.745 |
| 100 | 100 | 6.056 | 8.682 | 9.739 | 10.639 | 8.258 |

From the calculated values of reactivities presented in Tables 1 and 2 respectively, it is quite evident that the extent of enhancement in the total fusion reactivity due to suprathermal tail of the Dagum speed distribution, relative to that computed using the Maxwellian distribution, depends significantly on the values of the Dagum shaping parameters as well as the average ion kinetic temperature.

It can be clearly seen from Table 1, that at the bulk kinetic temperatures of 10, 15, 25 and 40 keV respectively, for a fixed value of shaping parameter 'a', the total fusion reactivity increases as the value of the parameter 'p' increases, indicating a significant enhancement in the volumetric reaction rate due to the enhancement in supra-thermal population. At an average kinetic temperature of 10 keV, with parameter 'a' and 'p' having fixed values of 3.0 and 5.0 respectively, an enhancement of about a factor 2.6 is observed for D-T total fusion reactivity, relative to the Maxwellian case at the same bulk temperature. The enhancement increases to about a factor of 3.76, when the value of parameter 'p' is increased to 8.0, with all other parameters remaining unchanged. For a bulk ion temperature of 15 keV and the aforesaid values of the Dagum shaping parameters, the enhancements in the total reactivity are about a factor of 1.7 and 2.3 respectively. At 25 keV, for the same parameter values the enhancements are calculated to be about a factor of 1.28 and 1.55 respectively. Thus, it is reasonable to conclude that, in the temperature ranges around and a little higher than the break- even temperature, the extent of enhancement in the total fusion reactivity decreases, as the average ion kinetic temperature increases. At a higher bulk kinetic temperature of 60 keV or above, an opposite trend is observed. Calculations for D-T plasma at 60, 80 and 100 keV show a steady decrease in the total fusion reactivity with the increase in the value of parameter 'p' for a fixed value of parameter 'a'. Although at 60 keV, for p= 5.0, some enhancement in reactivity is found relative to the Maxwellian case, it completely disappears at further higher temperatures. In fact at 80 keV and 100 keV, the Maxwellian speed distribution gives higher values of fusion reactivity compared to the values predicted by the Dagum counterpart.This change in the trend is attributed to the combined effect of fusion cross-section profile for the D-T fusion reaction [14] and the shape of ion speed distribution at higher kinetic temperatures, as shown in the earlier section.

Table 2 presents the total fusion reactivities for a fixed value of parameter 'p' and multiple values of parameter 'a'. Computational results show that at the bulk kinetic temperatures of 10, 15, 25 keV respectively, for a fixed value of shaping parameter 'p', the total fusion reactivity decreases as the value of the parameter 'a' increases. This trend is consistent with the nature of tail-enhancement found in the earlier section with the change in the value of shaping parameter 'a'.  At 10 keV, a significant increment in the volumetric reaction rate is observed due to the enhancement in supra-thermal population. For a 10 keV D-T plasma, with Dagum shaping parameters 'a' and 'p' having fixed values of 2.5 and 5.0 respectively, an enhancement of about a factor 3.54 is observed over the Maxwellian case .The enhancement disappears, when the value of parameter 'a' is increased to 4.5, with all other parameters remaining unchanged. At the kinetic temperature of 15 keV, for a= 2.5, an enhancement of about a factor of 2.03 over the Maxwellian case has been found. At 25 keV, for a= 2.5, an enhancement of about a factor 1.32 is observed. Thus, it is reasonable to conclude that in the temperature ranges around and a little higher than the break- even temperature, the extent of enhancement in the total fusion reactivity decreases as the average ion kinetic temperature increases, and it disappears with higher values of a. At a higher bulk kinetic temperature of 60 keV or above, an opposite trend is observed. Calculations for D-T plasma at 60, 80 and 100 keV show a steady increase in the total fusion reactivity with the increase in the value of parameter 'a' for a fixed value of parameter 'p'. At lower values of 'a', there is no enhancement in the total reactivity, but for a= 4.5, an enhancement of approximately a factor of 1.28 in total reactivity is observed in the temperature range 60-100 keV over the Maxwellian case. Thus it is quite evident that the ion speed distribution together with the fusion cross section profile determines the fusion reaction rate and the computational results provides us with an idea about the choice of operating temperature for optimum fusion power density. But, to decide on the operating temperature Lawson criterion has to be fulfilled and detailed power balance calculations have to be performed [1, 2].

## b. Comparison based on Equivalent Kinetic Temperature

For the comparison between the reactivities calculated using equivalent kinetic temperature approach, a set of kinetic temperatures for the Dagum distribution ($T_{Dg}$) is considered to begin with and then equivalent temperatures for Maxwellian distribution ($T_{Mx}$) are computed for each of the chosen values of $T_{Dg}$. As mentioned before, the equivalence between the kinetic temperatures belonging to two different ion speed distributions is obtained by equating the respective mean thermal speeds. In general the mean of 3-parameter Dagum speed distribution ($M_{Dg}$) [12] and that of Maxwellian distribution ($M_{Mx}$) [15] are expressed in the following way:

$$M_{Dg} = \left(\frac{-b_{Dg}}{a}\right)\left(\frac{\Gamma\left(1-\frac{1}{a}\right)\Gamma\left(\frac{1}{a}+p\right)}{\Gamma(p)}\right), \text{ (a>1, p>0); and } M_{Mx} = 2b_{Mx}\sqrt{\frac{2}{\pi}}$$

Where $b_{Mx} = \sqrt{\frac{k_B T_{Mx}}{\mu}}$ and the other parameters $a, b_{Dg}, p$ are as defined earlier. Now, equating $M_{Dg}$ and $M_{Mx}$, it can be shown that

$$T_{Mx} = T_{Dg}\left[\left(\frac{\pi}{8}\right)\left(\frac{\Gamma\left(1-\frac{1}{a}\right)\Gamma\left(\frac{1}{a}+p\right)}{\Gamma(p)}\right)^2\right] \tag{3}$$

Plasma kinetic temperatures of 10, 15, 25, 40, 60, 80 and 100 keV have been chosen respectively, and are considered as the background bulk temperatures for the ion population following the Dagum speed distribution. The equivalent Maxwellian temperatures corresponding to the above mentioned Dagum temperatures are calculated using Eq. 3. In this current work, equivalent kinetic temperatures have been calculated for multiple values of parameter 'p' (p= 5.0, 6.0, 7.0 and 8.0), keeping the value of parameter 'a' fixed (a= 3.0); as well as for multiple values of parameter 'a' (a= 2.5, 3.5, 4.0 and 4.5), while the parameter 'p' is kept fixed (p= 5.0). Thereafter the total reactivities are calculated using Eq. 1. Tables 3 to 6 present the total fusion reactivities for the fixed value of 'a', while the value of 'p' changes from one table to another. Total fusion reactivities for a fixed value of 'p', and changing values of parameter 'a' are presented in Tables 7 to 10, respectively.

**Table 3:** $\langle\sigma v\rangle_{DT,Dg}(a = 3.0, p = 5.0)$ and $\langle\sigma v\rangle_{DT,Mx}$ for equivalent background plasma kinetic temperature

| $T_{Dg}(keV)$ | $T_{Mx}(keV)$ | $\langle\sigma v\rangle_{DT,Dg}$ $* 10^{-22} \left(m^3/s\right)$ | $\langle\sigma v\rangle_{DT,Mx}$ $* 10^{-22} \left(m^3/s\right)$ |
|---|---|---|---|
| 10 | 20.131 | 3.024 | 4.441 |
| 15 | 30.197 | 4.734 | 6.799 |
| 25 | 50.327 | 7.327 | 8.690 |
| 40 | 80.524 | 9.173 | 8.735 |
| 60 | 120.786 | 9.364 | 7.652 |
| 80 | 161.048 | 8.527 | 6.508 |
| 100 | 201.310 | 7.453 | 5.543 |

**Table 4:** $\langle\sigma v\rangle_{DT,Dg}(a = 3.0, p = 6.0)$ and $\langle\sigma v\rangle_{DT,Mx}$ for equivalent background plasma kinetic temperature

| $T_{Dg}(keV)$ | $T_{Mx}(keV)$ | $\langle\sigma v\rangle_{DT,Dg}$ $*10^{-22}\left(m^3/s\right)$ | $\langle\sigma v\rangle_{DT,Mx}$ $*10^{-22}\left(m^3/s\right)$ |
|---|---|---|---|
| 10 | 22.905 | 3.504 | 5.217 |
| 15 | 34.357 | 5.372 | 7.436 |
| 25 | 57.262 | 7.989 | 8.866 |
| 40 | 91.619 | 9.493 | 8.483 |
| 60 | 137.428 | 9.155 | 7.163 |
| 80 | 183.237 | 7.978 | 5.952 |
| 100 | 229.046 | 6.736 | 4.988 |

**Table 5:** $\langle\sigma v\rangle_{DT,Dg}(a = 3.0, p = 7.0)$ and $\langle\sigma v\rangle_{DT,Mx}$ for equivalent background plasma kinetic temperature

| $T_{Dg}(keV)$ | $T_{Mx}(keV)$ | $\langle\sigma v\rangle_{DT,Dg}$ $*10^{-22}\left(m^3/s\right)$ | $\langle\sigma v\rangle_{DT,Mx}$ $*10^{-22}\left(m^3/s\right)$ |
|---|---|---|---|
| 10 | 25.520 | 3.951 | 5.859 |
| 15 | 38.280 | 5.938 | 7.895 |
| 25 | 63.800 | 8.508 | 8.916 |
| 40 | 102.081 | 9.362 | 8.200 |
| 60 | 153.122 | 8.833 | 6.721 |
| 80 | 204.162 | 7.412 | 5.483 |
| 100 | 255.203 | 6.080 | 4.536 |

**Table 6:** $\langle\sigma v\rangle_{DT,Dg}(a = 3.0, p = 8.0)$ and $\langle\sigma v\rangle_{DT,Mx}$ for equivalent background plasma kinetic temperature

| $T_{Dg}(keV)$ | $T_{Mx}(keV)$ | $\langle\sigma v\rangle_{DT,Dg}$ $*10^{-22}\left(m^3/s\right)$ | $\langle\sigma v\rangle_{DT,Mx}$ $*10^{-22}\left(m^3/s\right)$ |
|---|---|---|---|
| 10 | 28.009 | 4.368 | 6.391 |
| 15 | 42.013 | 6.441 | 8.228 |
| 25 | 70.022 | 8.909 | 8.889 |
| 40 | 112.035 | 9.648 | 7.911 |
| 60 | 168.052 | 8.457 | 6.326 |
| 80 | 224.069 | 6.869 | 5.082 |
| 100 | 280.087 | 5.499 | 4.161 |

**Table 7:** $\langle\sigma v\rangle_{DT,Dg}$ ($a = 2.5, p = 5.0$) and $\langle\sigma v\rangle_{DT,Mx}$ for equivalent background plasma kinetic temperature

| $T_{Dg}(keV)$ | $T_{Mx}(keV)$ | $\langle\sigma v\rangle_{DT,Dg}$ $*10^{-22} (m^3/s)$ | $\langle\sigma v\rangle_{DT,Mx}$ $*10^{-22} (m^3/s)$ |
|---|---|---|---|
| 10 | 30.074 | 4.106 | 6.777 |
| 15 | 45.110 | 5.671 | 8.439 |
| 25 | 75.184 | 7.556 | 8.827 |
| 40 | 120.295 | 8.419 | 7.666 |
| 60 | 180.442 | 7.989 | 6.019 |
| 80 | 240.589 | 7.039 | 4.781 |
| 100 | 300.737 | 6.056 | 3.885 |

**Table 8:** $\langle\sigma v\rangle_{DT,Dg}$ ($a = 3.5, p = 5.0$) and $\langle\sigma v\rangle_{DT,Mx}$ for equivalent background plasma kinetic temperature

| $T_{Dg}(keV)$ | $T_{Mx}(keV)$ | $\langle\sigma v\rangle_{DT,Dg}$ $*10^{-22} (m^3/s)$ | $\langle\sigma v\rangle_{DT,Mx}$ $*10^{-22} (m^3/s)$ |
|---|---|---|---|
| 10 | 15.389 | 2.124 | 2.922 |
| 15 | 23.084 | 3.745 | 5.264 |
| 25 | 38.473 | 6.697 | 7.915 |
| 40 | 61.556 | 9.419 | 8.909 |
| 60 | 92.335 | 10.363 | 8.465 |
| 80 | 123.113 | 9.769 | 7.583 |
| 100 | 153.891 | 8.682 | 6.700 |

**Table 9:** $\langle\sigma v\rangle_{DT,Dg}$ ($a = 4.0, p = 5.0$) and $\langle\sigma v\rangle_{DT,Mx}$ for equivalent background plasma kinetic temperature

| $T_{Dg}(keV)$ | $T_{Mx}(keV)$ | $\langle\sigma v\rangle_{DT,Dg}$ $*10^{-22} (m^3/s)$ | $\langle\sigma v\rangle_{DT,Mx}$ $*10^{-22} (m^3/s)$ |
|---|---|---|---|
| 10 | 12.693 | 1.459 | 2.012 |
| 15 | 19.040 | 2.877 | 4.111 |
| 25 | 31.733 | 5.905 | 7.054 |
| 40 | 50.773 | 9.303 | 8.706 |
| 60 | 76.160 | 11.037 | 8.812 |
| 80 | 101.547 | 10.777 | 8.215 |
| 100 | 126.934 | 9.739 | 7.469 |

**Table 10:** $\langle\sigma v\rangle_{DT,Dg}$ $(a = 4.5, p = 5.0)$ and $\langle\sigma v\rangle_{DT,Mx}$ for equivalent background plasma kinetic temperature

| $T_{Dg}(keV)$ | $T_{Mx}(keV)$ | $\langle\sigma v\rangle_{DT,Dg}$ $*10^{-22} \left(m^3/s\right)$ | $\langle\sigma v\rangle_{DT,Mx}$ $*10^{-22} \left(m^3/s\right)$ |
|---|---|---|---|
| 10 | 10.981 | 0.994 | 1.458 |
| 15 | 16.472 | 2.181 | 3.285 |
| 25 | 27.453 | 5.097 | 6.279 |
| 40 | 43.924 | 8.955 | 8.364 |
| 60 | 65.887 | 11.448 | 8.914 |
| 80 | 87.849 | 11.580 | 8.576 |
| 100 | 109.811 | 10.639 | 7.976 |

As can be seen from Tables 3 to 6, if the total reactivities calculated at the equivalent kinetic temperatures pertaining to the each of the ion speed distributions are compared against each other, Maxwellian distribution predicts higher total reactivities when compared against the values obtained using the Dagum distribution, at the Dagum kinetic temperatures of 10, 15 and 25 keV. At Dagum temperatures of 40 keV or higher, adequate enhancement in the total fusion reactivity is observed over the corresponding Maxwellian cases. At Dagum kinetic temperatures of 40 keV and 60 keV for a fixed value of Dagum parameter 'a' (a= 3.0), the extent of enhancement in total reactivity increases with an increase in the value of parameter 'p'. For p= 5.0 an enhancement of a factor of about 1.05 is observed at 40 keV and that amounts to a factor of about 1.22 at 60 keV over the Maxwellian counterpart. For p= 8.0, the enhancements at the same Dagum kinetic temperatures are found to be a factor of 1.22 and 1.34, respectively. At the higher Dagum temperatures of 80 and 100 keV, the extent of enhancement in the total reactivity calculated has been found to be almost uniform and amounts to a factor of about 1.3, for a= 3.0, irrespective of the value of parameter 'p'.

    Tables 7 to 10 show that Maxwellian distribution predicts higher total reactivities when compared against the values obtained using the Dagum distribution, at the Dagum kinetic temperatures of 10, 15 and 25 keV respectively irrespective of the values of Dagum shaping parameter 'a'. At Dagum temperatures of 40 keV or higher, adequate enhancement in the total fusion reactivity is observed over the corresponding Maxwellian cases. At Dagum kinetic temperatures of 40 keV and 60 keV for a fixed value of Dagum parameter 'p' (p= 5.0), the extent of enhancement in total reactivity decreases as the value of parameter 'a' increases. This is quite expected because as the value of parameter 'a' is increased, the supra-thermal tail gets depleted. For a= 2.5 an enhancement of a factor of about 1.10 is observed at 40 keV and that amounts to a factor of about 1.33 at 60 keV over the Maxwellian counterpart. For a= 4.5, the enhancements at the same Dagum kinetic temperatures are found to be a factor of 1.07 and 1.28, respectively. For lower values of parameter 'a' (a= 2.5), the extent of enhancement in total reactivity at even higher temperatures of 80 keV and 100 keV are about a factor of 1.47 and 1.56 respectively, but as the value of 'a' increases, the enhancement diminishes to a factor

of about 1.3. This is due to the combined effect of the shape of the ion speed distribution and the fusion cross-section profile pertinent to D-T reaction, as mentioned earlier.

## B. Comparison of Thermal and Supra-thermal Range Reactivities

Apart from comparing the total fusion reactivities, it is also of interest to compare both supra-thermal and thermal range fusion reaction rates calculated using the Dagum distribution with the respective Maxwellian case counterparts. Since Maxwellian is well known for describing the thermalized quasi-equilibrium scenarios in the magnetically confined fusion reactors and it has a more pronounced peak in the thermal range and Dagum distribution has been introduced in the current work, with a view to capturing the effect of the substantial supra-thermal accelerated ion population; the thermal speed and the supra-thermal speed ranges have been defined by the intervals $(0, m_{2/3})$ and $(d_{2/3}, \infty)$ respectively; where $m_{2/3}$ and $d_{2/3}$ are defined by the following integrals:

$$\int_0^{m_{2/3}} f_{Mx}(v) dv = 2/3, \quad \text{and} \quad \int_{d_{2/3}}^{\infty} f_{Dg}(v) dv = 1/3$$

$f_{Mx}(v)$ & $f_{Dg}(v)$ being the density of Maxwellian and Dagum speed distributions, respectively.

The thermal range cut-off speed $m_{2/3}$ can be found in the following way:
If $v$ represents the ion speed in the Maxwellian speed distribution, and the scale parameter is given by $b_{Mx} = \sqrt{\frac{k_B T_{Mx}}{\mu}}$, then the quantity $\left(\frac{v}{b_{Mx}}\right)$ follows the Standard Maxwellian distribution and hence $\left(\frac{v^2}{b_M^2}\right)$ follows Chi-Square distribution with degrees of freedom 3 (i.e. $\chi_3^2$ distribution). Now by the definition of $m_{2/3}$, it is obvious that $P(v \leq m_{2/3}) = 2/3$.

Using the transformation between the $\chi^2$ and Maxwellian distribution, it can be shown that

$$P\left(\chi_3^2 \leq \frac{m_{2/3}^2}{b_{Mx}^2}\right) = 2/3.$$

Using tables available for $\chi^2$ distribution **[16]**, it can be found that -

$$m_{2/3} \approx b_{Mx}\sqrt{3.404706}. \tag{4}$$

The supra-thermal range lower cut-off speed $d_{2/3}$ can be calculated using the cumulative distribution function (CDF) of the Dagum distribution $Dg(a, p, b)$.

The CDF for $Da(a, p, b)$ is defined as follows [9]:

$$F(x) = \left(1 + \left(\frac{x}{b_{Dg}}\right)^{-a}\right)^{-p} \tag{5}$$

where, a, p and $b_{Dg}$ bear the same meanings as mentioned before.

By the definition of $d_{2/3}$, it is obvious that $F(d_{2/3}) = 2/3$ and hence by a simple calculation, it can be shown that $d_{2/3} = b_{Dg}\left[\left(\frac{3}{2}\right)^{\frac{1}{p}} - 1\right]^{-1/a}$.

    With the thermal and supra-thermal cut-off speeds being defined in terms of the bulk temperature and distribution –shaping parameters; two new ratios have been defined for the quantitative comparison between the thermal range and supra-thermal range reactivities predicted by the two different ion speed distributions of interest.

    The first ratio ($R_1$) is termed as the 'Thermal Reactivity Ratio' and is defined mathematically as follows:

$$R_1 = \frac{\int_0^{m_{2/3}} v f_{Dg}(v) \sigma(v) dv}{\int_0^{m_{2/3}} v f_{Mx}(v) \sigma(v) dv} \tag{6}$$

The second ratio ($R_2$) of interest is termed as the 'Suprathermal Reactivity Ratio' and is defined by the following mathematical expression:

$$R_2 = \frac{\int_{d_{2/3}}^{\infty} v f_{Dg}(v) \sigma(v) dv}{\int_{d_{2/3}}^{\infty} v f_{Mx}(v) \sigma(v) dv} \tag{7}$$

Using Eqs. (6) and (7) the reactivity ratios are calculated and is expressed in terms of the Dagum shaping parameters 'a' and 'p' in the following subsection.

a. **Variation in Thermal and Supra-thermal Contributions based on Shaping Parameters**

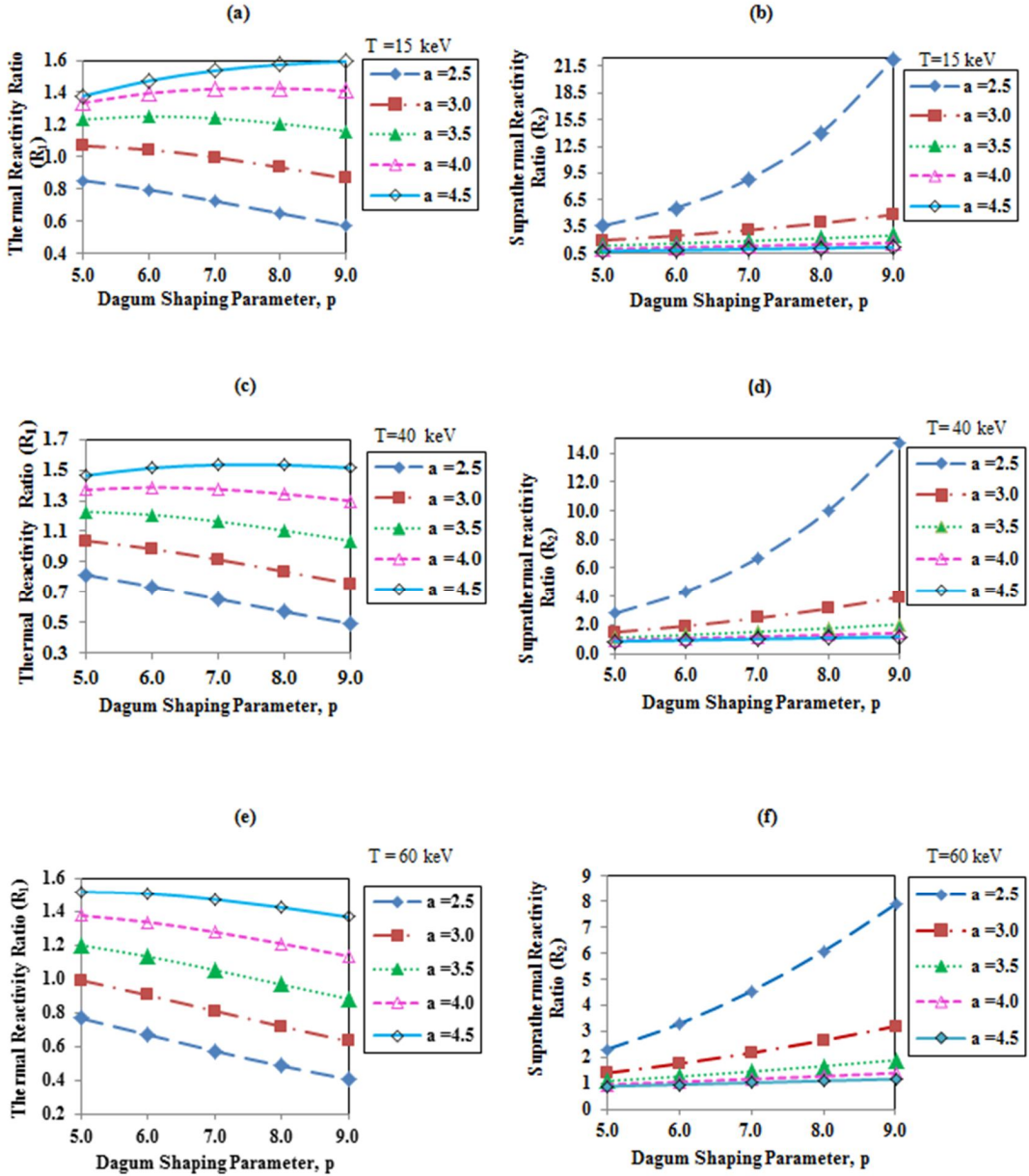

**Fig. 4** Panel **a.** $R_1$ at 15 keV, **b.** $R_2$ at 15 keV, **c.** $R_1$ at 40 keV, **d.** $R_2$ at 40 keV, **e.** $R_1$ at 60 keV, and **f.** $R_2$ at 60 keV versus Dagum Shaping Parameter (p) plot for multiple values of Shaping Parameter 'a'.

Upon calculating $R_1$ and $R_2$, the values of interest are plotted in Figure 4. Since, numerical integration has been used to find the thermal and supra-thermal fusion reactivities pertaining to both Maxwellian and Dagum speed distributions, thus the calculations of the ratios $R_1$ and $R_2$ contain some inherent errors. With a view to estimating the reactivity ratios in a reasonably accurate fashion, as a function of the Dagum shaping parameters; a polynomial regression, also called as response surface, has been performed on parameters $p\ and\ a$, for a fixed background kinetic temperature. From the Figure 4, it can be seen that both the reactivity ratios show a second order nature, when plotted against the Dagum shaping parameters. Thus, in the current research, optimal regression surfaces are obtained for each of $R_1$ and $R_2$ for multiple background kinetic temperatures 15, 40, 60 and 80 keV respectively; on the basis of the dataset {( $R_1, R_2, T, p, a$): T = 15, 40, 60, 80 keV ; p = 5.0, 6.0, 7.0, 8.0, 9.0 ; a= 2.5, 3.0, 3.5, 4.0, 4.5}, among all the subsets of the second order regression surfaces and they are given by following mathematical expressions:

$$R_i = \alpha + \beta_1 p + \beta_2 a + \gamma_1 p^2 + \gamma_2 a^2 + \gamma_3 pa; \quad (i = 1, 2) \tag{8}$$

Optimal values of the model parameter set $(\alpha, \beta_1, \beta_2, \gamma_1, \gamma_2, \gamma_3)$ for each of $R_1$ and $R_2$, for each of the different kinetic temperatures have been found by systematically using the following steps:

The best model is chosen based on two 'goodness of fit' criteria, namely adjusted $R^2$ and Mallow's $C_p$ **[17]**. If different models are suggested by these two criteria, then the intersection of those two models is considered as the final model and is fitted to the relevant dataset. Afterthat, the model is again reduced by looking at the 'p-values' **[18, 19],** corresponding to the intercept and the coefficients. This reduction is done term by term, by starting with the intercept term and then moving to the linear term, followed by the quadratic term and the interaction term. Each of the p-values corresponding to the coefficients in the final model has been chosen to be strictly less than 0.01. The level of significance has been considered to be such a small value in order to have the proposed surfaces as simple as possible in each of the cases.

In order to fit a regression surface, three basic assumptions need to be checked regarding the associated errors, namely "Independence", "Homoscedasticity across the predictors" and "Normality" **[20, 21]**. Now, "Independence" of the errors is inherent in the present scenario since, numerical integration is carried out separately for each of the choices of the set $(T, p, a)$. The assumption of "Homoscedasticity" is checked statistically by employing Levene's test **[22]** with respect to median to the residuals across 'p' and 'a'. The assumption of "Normality" is examined statistically by utilizing Shapiro Wilk test **[23]**. The models found using the aforesaid 'goodness of fit' and 'p-value' criteria have been found to satisfy the assumptions of "Homoscedasticity" and "Normality" of errors statistically, showing p-values greater than 0.1 and 0.09 respectively, and hence the fitting of the proposed regression surfaces are justified. Also, the coefficient of determination ($R^2$) for each of the regression surfaces has been found to be close to 1. It is to be mentioned that the obtained regression surfaces can be interpolated within the considered dataset of Dagum shaping parameters and the average kinetic temperature and cannot be extrapolated beyond. The optimal values of the intercept and other coefficients for the regression surfaces of interest are listed in Table 11.

**Table 11:** Optimal Choices for the Regression Surface Parameters and the associated $R^2$ values

| Fractional contributions | $T$ (keV) | Optimal choices of the model parameters | | | | | | Coefficient of Determination $R^2$ |
|---|---|---|---|---|---|---|---|---|
| | | $\alpha$ | $\beta_1$ | $\beta_2$ | $\gamma_1$ | $\gamma_2$ | $\gamma_3$ | |
| $R_1$ | 15 | 0 | −0.1368 | 0.6060 | −0.0077 | −0.0954 | 0.0658 | 0.9999 |
| | 40 | 0 | −0.1169 | 0.5492 | −0.0066 | −0.0648 | 0.0479 | 0.9999 |
| | 60 | 0 | −0.1458 | 0.5923 | 0 | −0.0425 | 0.0213 | 0.9998 |
| | 80 | 5.0525 | −0.0926 | 0.4867 | 0 | −0.0103 | 0 | 0.9998 |
| $R_2$ | 15 | 0 | 8.7780 | −14.6544 | 0 | 3.6447 | −2.1614 | 0.8926 |
| | 40 | 0 | 5.9203 | −9.6847 | 0 | 2.4250 | −1.4555 | 0.9219 |
| | 60 | 0 | 3.0650 | −4.4756 | 0 | 1.1395 | −0.7422 | 0.9666 |
| | 80 | 0 | 1.3185 | −4.2716 | 0 | 0.7619 | −0.2987 | 0.9671 |

**Conclusion**

From the results presented in the current work, it can be concluded that the extent of enhancement in the total fusion reactivity pertaining to different relevant fusion reactions depends on the shape of the speed distribution as well as the energy dependent fusion cross-section profile. The shape of a speed distribution function depends on the associated shape and scale parameters. In case of Dagum speed distribution, the scale parameter includes the average ion kinetic temperature. Thus the bulk kinetic temperature of the background plasma becomes important in determining the enhancement in the total fusion reactivities. In the equivalent temperature approach of comparing the total fusion reactivities, instead of equating the thermal mean speeds, the median speeds from each of the distributions can be equated. The 'median-speed-equivalence' approach provides with a greater flexibility because median is explicitly defined for any distribution irrespective of the existence of the moments and thus no restriction is needed on the distribution parameter space for finding median. Additionally, regression surfaces found for the thermal and suprathermal reactivity ratios, respectively, provides us with a way to understand the extent of enhancement of reactivity due to Dagum distribution relative to the thermalized Maxwellian case, for different shape parameters as well as average kinetic temperature; without the information overhead about the fusion cross sections and actual shapes of the speed distributions.

**Acknowledgement**

The authors are grateful to Dr. Consuelo Arellano, faculty of the Statistics department at North Carolina State University, for her invaluable inputs in fitting the regression surfaces.